\documentclass[letter]{aa} 
\usepackage{graphicx}
\usepackage{txfonts}
\usepackage{natbib}
\usepackage{color}
\usepackage{amsmath}    
\usepackage{amssymb}    

\def\arcsec{"}

\def\aap{A\&A}

\def\apjl{ApJL}

\def\lesssim{\mathrel{\hbox{\rlap{\hbox{\lower4pt\hbox{$\sim$}}}\hbox{$<$}}}}
\def\gtrsim{\mathrel{\hbox{\rlap{\hbox{\lower4pt\hbox{$\sim$}}}\hbox{$>$}}}}

\def\bjdtdb{\ensuremath{\rm {BJD_{TDB}}}}
\def\oversim#1#2{\lower0.5pt\vbox{\baselineskip0pt \lineskip-0.5pt 
     \ialign{$\mathsurround0pt #1\hfil##\hfil$\crcr#2\crcr\sim\crcr}}} 

\begin{document} 

\title{ The $\delta$ Scuti pulsations of $\beta$ Pictoris\\
 as observed by ASTEP from Antarctica}
\subtitle{ }

\author{
D.~M\'ekarnia\inst{1} \thanks{E-mail: mekarnia@oca.eu}, E.~Chapellier\inst{1}, T.~Guillot\inst{1}, L.~Abe\inst{1},  
A.~Agabi\inst{1}, Y.~De~Pra\inst{2}, F.-X.~Schmider\inst{1}, K.~Zwintz\inst{3}, K.~B.~Stevenson\inst{4}, J.~J.~Wang\inst{5}, A.-M.~Lagrange\inst{6}, 
L.~Bigot\inst{1}, N.~Crouzet\inst{7}, Y.~Fante\"\i-Caujolle\inst{1}, J.-M.~Christille\inst{8} and P.~Kalas\inst{5,9}
}
\institute{
Universit\'{e} C\^ote d'Azur, Observatoire de la 
 C\^ote d'Azur, CNRS, Laboratoire Lagrange, CS 34229, F-06304 Nice Cedex 4, France 
\and 
Concordia Station, Dome\,C, Antarctica 
\and 
Institut f\"ur Astro- und Teilchenphysik, Universit\"at Innsbruck, Technikerstrasse 25/8, 6020 Innsbruck, Austria
\and
Space Telescope Science Institute, Baltimore, MD 21218, USA
\and 
Astronomy Department, University of California, Berkeley, Berkeley CA 94720, USA
\and 
Universit\'e Grenoble Alpes, CNRS, IPAG, 38\,000 Grenoble, France
\and 
Instituto de Astrofisica de Canarias, C/Via Lactea s/n, E-38200 La Laguna, Spain
\and
Fondazione Cl\'ement Fillietroz ONLUS, Astron. Observatory of the Autonomous region Aosta Valley, NUS, 11020 (AO), Italy
\and
SETI Institute, Carl Sagan Center, 189 Bernardo Avenue, Mountain View, CA 94043, USA
}

\date{Received YYY; Accepted XXX}
\titlerunning{ }
\authorrunning{ }

 \abstract
  {} 
  {The Antartica Search for Transiting Extrasolar Planets (ASTEP), an automatized 400 mm telescope located at Concordia station in Antarctica, monitored $\beta$~Pictoris continuously to detect 
any variability linked to the transit of the Hill sphere of its planet $\beta$~Pictoris~b. The long observation sequence, from March to 
September 2017, combined with the quality and high level duty cycle of our data, enables us to detect and analyse the $\delta$ Scuti pulsations of the star.}
  {Time series photometric data were obtained using aperture photometry by telescope defocussing. The 66\,418 data points were analysed using the 
software package Period04. We only selected frequencies with amplitudes that exceed four times the local noise level in the amplitude spectrum.}
  {We detect 31 $\delta$ Scuti pulsation frequencies, 28 of which are new detections. All the frequencies detected are in the interval 
34.76$-$75.68\,d$^{-1}$. We also find that $\beta$~Pictoris exhibits at least one pulsation mode that varies in amplitude over our monitoring 
duration of seven months.}
  {}

\maketitle


\section{Introduction}

With a visual magnitude of 3.8, the A5V star $\beta$~Pictoris is among the 500 brightest stars in the sky. This star is surrounded by a debris disk composed 
of dust and gas known to be continuously replenished by evaporating exocomets and colliding 
planetesimals (\citealt{Ferlet+1987}; \citealt{Lecavelier+1996}; \citealt{Beust+2000}; \citealt{Wilson+2017}), and hosts at least one planet, $\beta$ Pictoris b, detected 
with the Very Large Telescope's (VLT) adaptive optics NaCo system (\citealt{Lagrange+2009}; \citealt{Lagrange+2010}). The disk, which is  
seen edge-on, offers a unique opportunity to study the planet-forming environment and planet-disk interactions.

Many studies have been devoted to the physical characterization of the environment of the star. Studies of $\beta$~Pictoris itself have 
mainly focussed on the determination of its basic parameters and age (\citealt{Lanz+1995}; \citealt{Crifo+1997}; 
\citealt{DiFolco+2004}). \citealt{Crifo+1997} found that the star is very close to the zero age main sequence (ZAMS), which is a finding 
confirmed by studies of the age of the $\beta$~Pictoris moving group, 23$\pm 3$\,Myr \citep{Mamajek+Bell2014}. 

Very low-amplitude periodic variations in brightness, radial velocity, and line profile have been reported and 
studied (\citealt{Koen+2003a}; \citealt{Koen+2003b}; \citealt{Galland+2006}). These observations showed that $\beta$~Pictoris is one of the brightest 
known $\delta$~Scuti stars. Using photometric time series with a 50 cm telescope, \citealt{Koen+2003a} identified three frequencies with amplitudes 
below 1.5\,mmag, while 18 frequencies were identified from line-profile spectroscopic variations obtained with the SAAO 1.9 m telescope \citep{Koen+2003b}.

Since its discovery in 2003, the monitoring of the astrometric position of the planet $\beta$~Pictoris~b has provided a good estimation of 
its orbital parameters. It appears that the orbit of $\beta$~Pictoris is very close to the transit configuration. As a consequence, it is certain 
that the Hill sphere of $\beta$~Pictoris~b transits the star with a closest approach either in August 2017 \citep{Wang+2016}, October 
2017 (Lagrange, in preparation), or  January 2018  \citep{LecavelierDesEtangs+2016}. This has motivated a  monitoring campaign, started in 
2017 and still ongoing, to detect any variability of $\beta$~Pictoris due to circumplanetary material surrounding its planet (see e.g. \citealt{Stuick+2017}).

In this Letter, we analyse  a long-sequence photometric observation, part of the monitoring campaign, of $\beta$~Pictoris. These observations 
 allow us to better characterize the pulsations of the star. Section\,\ref{sect:observations} of the paper describes the observations 
and data reduction while Section\,\ref{sect:period} focusses on the frequency analysis. Results are discussed in Section\,\ref{sect:discussion} and 
concluding remarks are made in Section\,\ref{sect:conclusion}.

\section{Observations and data reduction}
\label{sect:observations}

Photometric observations were made with the Antartica Search for Transiting Extrasolar Planets (ASTEP) 400 mm telescope, installed at the Concordia station, Dome\,C,  
Antarctica (\citealt{Abe+2013}; \citealt{Guillot+2015}; \citealt{Mekarnia+2016}). Briefly, ASTEP\,400 is a custom 400 mm Newtonian telescope equipped 
with a 5-lens Wynne coma corrector and a 4k\,$\times$\,4k front-illuminated FLI Proline KAF\,16801E CCD with a 16 bits analogue-to-digital 
converter. The corresponding field of view is  1$^{\circ}\times1^{\circ}$ with an angular resolution of 0.93\arcsec pixel$^{-1}$  
(see \citealt{Daban+2010} for details). 

$\beta$ Pictoris, which is circumpolar under Dome\,C latitude, was monitored continuously from the beginning of March through the end of September 
of the 2017 Antarctic polar campaign; data acquisition started automatically when the Sun was 6 degrees below the horizon. These observations 
are part of the international campaign dedicated to monitoring the $\beta$~Pictoris~b Hill sphere transit, providing the opportunity 
to study the circumplanetry material surrounding the planet. $\beta$~Pictoris is so bright ($M_v$=3.85\,mag) that it saturates our CCD camera even at 
its minimum exposure time. We therefore used a highly defocussed point spread function (PSF) of approximately 100 pixels in 
diameter and containing roughly $10^8$ counts from  $\beta$~Pictoris.   The PSF value was chosen as a compromise between field crowdedness and 
signal-to-noise ratio compatible with a milli-magnitude precision of each individual calibrated point (considering photon noise only). The $\beta$~Pictoris 
stellar field is not very crowded so that the defocussed PSFs of all our reference stars do not overlap with any other 
nearby stars, which are bright enough to be detected within the exposure time. The Sloan {\it i'} filter ($\lambda$\,=\,0.695\,-\,0.844\,$\mu$m) was used. 
We monitored the star continuously when the sky brightness due to the Sun elevation allowed it, at a cadence 
of $\sim$90\,s (exposure time of 60\,s, readout, and processing overhead of $\sim$30\,s). In parallel, the focussed PSF is tracked by an independent 
CCD (see \citealt{Daban+2010}) at a cadence of about 1\,s. Each acquired science image passes through a custom PSF identification and field recognition 
algorithm to guarantee the field centring relative to the tracking CCD: the tracking CCD set point is 
constantly adjusted to compensate small differential drifts between the science and tracking optical paths. 
Data acquisition was performed automatically. To  maximize the recording continuity, we used  a reduced exposure time of 30\,s when the Sun elevation is 
between $-6^{\circ}$ and $-8^{\circ}$ due to the increased sky brightness.  

\begin{figure}
\includegraphics[width=\linewidth]{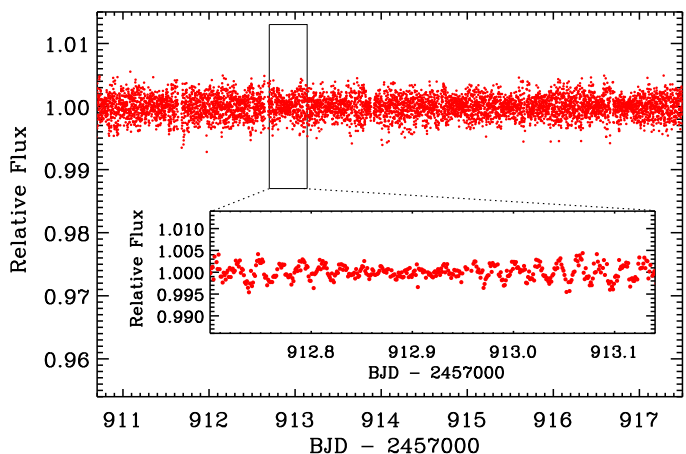} 
\caption{Seven-day light curve of $\beta$~Pictoris obtained between 7 and 13 June 2017, showing that data were 
continuously recorded during that period. The inset frame is a zoom-in on an area ($\sim$12-h) of the plot, showing 
the $\delta$~Scuti photometric variations of the star.} 
\label{fig:a_lc}
\end{figure}

The ASTEP\,400 data reduction pipeline includes bias and dark subtraction, bad pixel masking, and correction of the sky concentration 
effect, which causes an additive bell-shaped halo component to be present in the centre of each frame \citep{Andersen+1995}. 
We performed aperture photometry of the 18 brightest stars in our images with sufficient counts to be useful.  We used the same circular aperture 
radius for the 18 extracted sources, including $\beta$~Pictoris and our comparison stars HD\,39463 and HD\,38891. Figure\,\ref{fig:a_lc} shows a seven-day subset 
of the calibrated light curve obtained by ASTEP\,400. The rms noise is estimated at 1130\,ppm for an integration time of 60\,s.

The comparison data from the Hubble Space Telescope (HST) were obtained with WFC3/UVIS in the 953N 
filter ($\lambda$=0.9532\,$\mu$m, FWHM=0.0064\,$\mu$m) using spatial scanning mode to avoid saturation (HST programme GO-14621; PI: Wang). Photometry 
from each spatial scan was extracted using the WFC3 pipeline described in \citealt{Stevenson+2014}, turning each scan into a photometric point with 
uncertainties dominated by photon noise (57\,ppm). The HST data were then corrected for a 0.15\% photometric offset 
between the two scan directions by simultaneously fitting the offset with the $\delta$~Scuti pulsations as a 
quasi-periodic Gaussian process. A detailed analysis of the HST data will be presented in a future publication.

\section{Frequency analysis}
\label{sect:period}

\begin{figure}
\includegraphics[width=\linewidth]{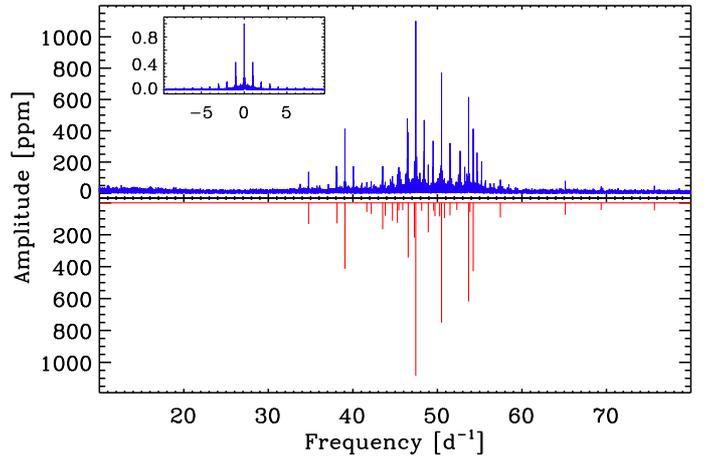} 
\caption{ Top panel: Amplitude spectrum of the cleaned photometric data of $\beta$~Pictoris. The spectral window is shown in 
the left corner of the top panel. Bottom panel: Amplitude spectrum with only the detected 31 frequencies resulting from the analysis process.} 
\label{fig:period}
\end{figure}

For our analysis, we kept a data set of 66\,418 good measurements (i.e. keeping only data with the Sun below -8$^{\circ}$ and selecting those 
obtained  under good photometric conditions) collected from March 5, to September 22, 2017, with an average sampling of 90\,s. This 
corresponds to 34.3\% of the total time, between the beginning and end of the observing period, and  52.7\% of the possible 
observing time (with the Sun below $-8^{\circ}$). The frequency analysis was performed with the software Period04 \citep{Lenz+2005}, which is based on classical 
Fourier analysis and least-square algorithms. We searched for frequencies in the interval 0\,-\,200\,d$^{-1}$, but all the $\delta$~Scuti frequencies 
were detected in the spectral region 34.76\,-\,75.68\,d$^{-1}$. The relative amplitude of the daily side lobes 
is kept below 0.4 (Fig.\,\ref{fig:period}) owing to
our excellent time coverage, thereby avoiding any ambiguity when extracting each frequency peak.

For each detected frequency the amplitude and phase were calculated by a least-squares sine fit. The data were then pre-\-whitened and a new analysis 
was performed. Our analysis was conducted for all frequencies detected with a signal-to-noise amplitude ratio above the widely used 
value S/N\,$=$\,4 \citep{Breger+1993}.  We detected 31\,$\delta$~Scuti frequencies, which are presented in Table\,\ref{tab:tab_period} and illustrated in 
Fig.\,\ref{fig:period}. The top panel of the figure shows the amplitude spectrum of our data, including the spectral window based on the times of 
analysed data, while the bottom panel shows the location of the detected frequencies. We reached a residual noise level of 9.45\,ppm in 
the 35\,-\,65\,d$^{-1}$ region.

\begin{figure}
\includegraphics[width=\linewidth]{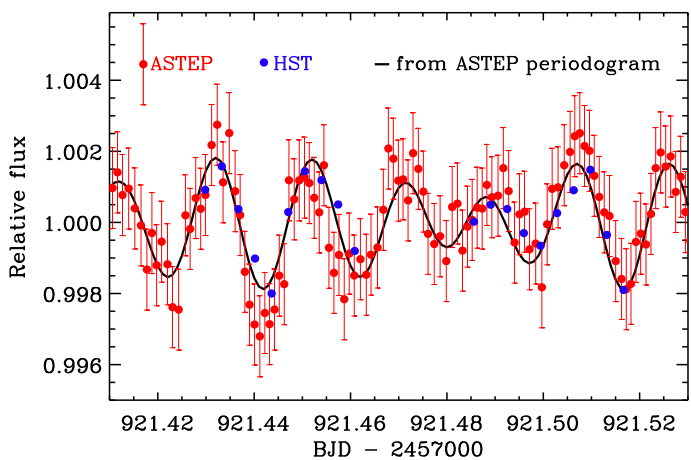} 
\caption{Comparison between ASTEP\,400 (red filled circles) and HST/WFC3 (blue filled circles) photometric measurements of $\beta$~Pictoris 
obtained on June 17, 2017. The 1\,$\sigma$ values are shown as error bars, which are too small to be seen for HST. The synthetic curve, calculated 
using our multiple-frequency analysis, is represented as a solid line.} 
\label{fig:previs}
\end{figure}

The uncertainties in the frequencies and their corresponding amplitudes and phases were computed with the relations proposed 
by \citet{Montgomery+1999}. However, as mentioned in their paper, these uncertainties are underestimates of the true size of the errors. Therefore, the 
uncertainties given in Table\,\ref{tab:tab_period} are 3$\sigma$ errors. We note that the relations proposed by \citet{Montgomery+1999} give the same 
error for all amplitudes. We also performed another independent check to investigate the reality of the detected frequencies. We created two equal 
subsets of the ASTEP time series and analysed these separately. We found a strong match between the frequencies obtained from the two subsets. The 
fit of the 31-frequency solution is plotted, as an example, over a small set of our measurements and 
better accuracy HST/WFC3 data points (Fig.\,\ref{fig:previs}). The fit is given by the 
relation $F(t) = 1 +\sum_{j=1}^{31} a_j \sin[2\pi f_j (t-t_0) + \phi_j]$, where $F(t)$ is the relative flux of the star as a function of the 
time and $f_j, a_j, \phi_j$ are the detected frequencies, amplitudes, and phases, respectively. The origin of the time 
is $t_0$(\bjdtdb)$=$2\,457\,000. Figure\,\ref{fig:previs} also highlights the complementarity between high accurate and stable HST and the much 
longer ASTEP data. The agreement is very good. The amplitude of the variations seems to be slightly lower 
in the HST data, than ASTEP data, which is expected because HST uses a redder filter (0.9532\,$\mu$m instead of 0.7695\,$\mu$m).

\begin{table}
\caption{\label{tab:tab_period} Thirty-one frequencies identified in the amplitude spectrum. Each value is presented with its 3$\sigma$ estimate.}
\begin{center}
\begin{tabular}{rrr}
    \hline
    \hline
Frequency ~~~~~~~~~       & Amplitude   & Phase ~~~~~~   \\
(d$^{-1}$)~~~~~~~~~~~~~                & (ppm)~~~       & (rad) ~~~~~~~   \\
\hline
 f$_{1}$ ~~  47.43904 $\pm$0.00005  &    1084 $\pm$19   &   1.994 $\pm$0.017\\
 f$_{2}$ ~~  50.49188 $\pm$0.00007  &     752 $\pm$19   &   4.777 $\pm$0.025\\
 f$_{3}$ ~~  53.69138 $\pm$0.00008  &     622 $\pm$19   &   1.022 $\pm$0.030\\
 f$_{4}$ ~~  54.23727 $\pm$0.00012  &     429 $\pm$19   &   1.477 $\pm$0.043\\
 f$_{5}$ ~~  39.06297 $\pm$0.00012  &     412 $\pm$19   &   2.767 $\pm$0.045\\
 f$_{6}$ ~~  46.54281 $\pm$0.00015  &     344 $\pm$19   &   3.359 $\pm$0.054\\
 f$_{7}$ ~~  47.28350 $\pm$0.00023  &     216 $\pm$19   &   0.077 $\pm$0.086\\
 f$_{8}$ ~~  48.91877 $\pm$0.00027  &     186 $\pm$19   &   0.620 $\pm$0.100\\
 f$_{9}$ ~~  43.52749 $\pm$0.00031  &     165 $\pm$19   &   3.433 $\pm$0.113\\
 f$_{10}$ ~  34.76028 $\pm$0.00038  &     133 $\pm$19   &   3.558 $\pm$0.141\\
 f$_{11}$ ~  38.12913 $\pm$0.00040  &     128 $\pm$19   &   1.683 $\pm$0.146\\
 f$_{12}$ ~  45.26910 $\pm$0.00041  &     125 $\pm$19   &   4.162 $\pm$0.149\\
 f$_{13}$ ~  44.68376 $\pm$0.00046  &     111 $\pm$19   &   0.424 $\pm$0.168\\
 f$_{14}$ ~  50.83133 $\pm$0.00053  &      95 $\pm$19   &   5.271 $\pm$0.196\\
 f$_{15}$ ~  57.45220 $\pm$0.00055  &      92 $\pm$19   &   4.186 $\pm$0.203\\
 f$_{16}$ ~  47.27074 $\pm$0.00056  &      91 $\pm$19   &   5.205 $\pm$0.205\\
 f$_{17}$ ~  50.26818 $\pm$0.00060  &      85 $\pm$19   &   1.093 $\pm$0.220\\
 f$_{18}$ ~  49.71224 $\pm$0.00063  &      81 $\pm$19   &   6.030 $\pm$0.231\\
 f$_{19}$ ~  43.82897 $\pm$0.00064  &      80 $\pm$19   &   3.344 $\pm$0.233\\
 f$_{20}$ ~  51.49786 $\pm$0.00066  &      77 $\pm$19   &   2.491 $\pm$0.242\\
 f$_{21}$ ~  65.13460 $\pm$0.00068  &      75 $\pm$19   &   0.938 $\pm$0.248\\
 f$_{22}$ ~  42.17268 $\pm$0.00074  &      69 $\pm$19   &   3.588 $\pm$0.271\\
 f$_{23}$ ~  53.85402 $\pm$0.00089  &      57 $\pm$19   &   2.518 $\pm$0.328\\
 f$_{24}$ ~  41.65042 $\pm$0.00093  &      55 $\pm$19   &   1.920 $\pm$0.340\\
 f$_{25}$ ~  49.55851 $\pm$0.00104  &      49 $\pm$19   &   3.919 $\pm$0.382\\
 f$_{26}$ ~  48.13791 $\pm$0.00106  &      48 $\pm$19   &   1.010 $\pm$0.390\\
 f$_{27}$ ~  45.43672 $\pm$0.00109  &      47 $\pm$19   &   2.957 $\pm$0.398\\
 f$_{28}$ ~  45.90177 $\pm$0.00113  &      45 $\pm$19   &   5.438 $\pm$0.414\\
 f$_{29}$ ~  75.67860 $\pm$0.00113  &      45 $\pm$19   &   3.357 $\pm$0.416\\
 f$_{30}$ ~  69.37515 $\pm$0.00116  &      44 $\pm$19   &   0.997 $\pm$0.426\\
 f$_{31}$ ~  52.30184 $\pm$0.00116  &      44 $\pm$19   &   5.801 $\pm$0.427\\
   \hline
\end{tabular}
\end{center}
\end{table}

\section{Discussion}
\label{sect:discussion}

All the detected frequencies are confined in the interval between  34.76 and 75.68\,d$^{-1}$. These frequencies are high for 
a $\delta$~Scuti, but not exceptional among hot pre-main-sequence (PMS) $\delta$~Scuti star (\citealt{Casey+2011}; \citealt{Zwintz+2014}). The 
non-regularity of the frequencies in the spectrum is also common for the PMS and ZAMS. 

We did not detect any $2f$ frequency. So, all the pulsation modes are sinusoidal, which is consistent with the small amplitudes we observed. We did not detect either $f_i$\,+\,$f_j$ combinations,  regular frequency spacing between consecutive eigenmodes, or regular rotational splitting. The rapid rotation of the 
star (vsini\,$=$\,130\,km\,s$^{-1}$; \citealt{Royer+2007}) certainly breaks any regular pattern in the spectrum (see e.g. \citealt{Reese+2017}).

\begin{figure}
\includegraphics[width=\linewidth]{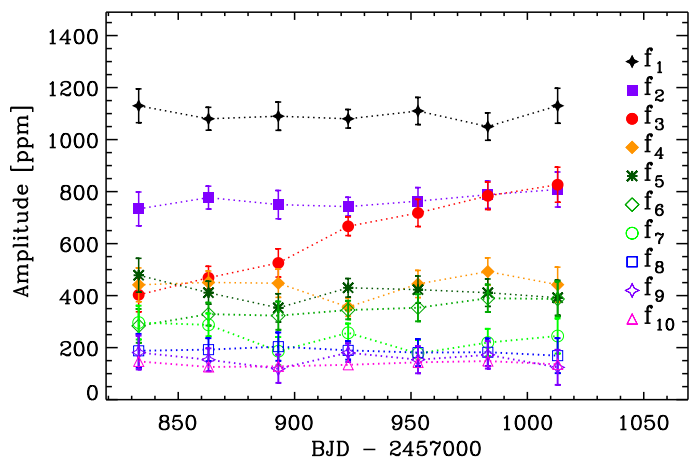} 
\caption{ Amplitude variations of the 10 main frequencies of $\beta$~Pictoris, showing the amplitude variation of 
third pulsation mode ($f_3$=53.69138\,d$^{-1}$). The 3\,$\sigma$ values 
are shown as error bars.}
\label{fig:ampvar}
\end{figure}

We looked for amplitude and phase modulation of the first 10 frequencies. We divided the data set into seven parts, each 30 days in length, and 
tracked the amplitude of our 10 highest amplitude peaks through the data set at fixed frequency using least squares. We found 
an amplitude change for only one pulsating mode that corresponds to our third pulsation frequency ($f_3$=53.69138\,d$^{-1}$) (Fig.\,\ref{fig:ampvar}). This 
amplitude increased from 403 to 826$\pm$66\,ppm. No significant amplitude variation was found for the other frequencies. On the other hand, no 
significant phase variation was detected in our analysis.  However, the detection of amplitude variations in our seven-month observations is not 
surprising since 61.3\%  of the $\delta$~Scuti stars observed by the Kepler Space Telescope exhibit at least one pulsation mode that varies 
significantly in amplitude over four years \citep{Bowman+2016}.

\label{sect:previs}

\cite{Koen+2003a} and  \cite{Koen+2003b} detected three frequencies (47.4355, 38.0593 and 47.2823\,d$^{-1}$) in 
their photometric and radial velocity observations with amplitudes $a$=1.52, 1.02, 0.88\,mmag, 
and $K$=0.129, 0.119, 0.03\,km\,s$^{-1}$, respectively. Their first and third frequencies correspond to our $f_1$=47.43904 and $f_7$=47.28350\,d$^{-1}$ frequencies 
but their 38.0593\,d$^{-1}$ frequency is clearly a one-day alias of our $f_5$=39.06297\,d$^{-1}$ frequency. \cite{Galland+2006} also 
detected two of the these frequencies  in their HARPS measurements $f$=47.44\,d$^{-1}$, $K$=0.215\,km\,s$^{-1}$ and $f$=39.05\,d$^{-1}$, $K$=0.17\,km\,s$^{-1}$.

 \cite{Koen+2003b} searched for periodicity in the moving bumps seen in the broaden line profile of the 
star. They detected 18 high-degree frequencies between 24.73 and 68.54\,d$^{-1}$. Most of these frequencies correspond to high-degree 
modes (4\,$\leq$\,$\vert m \vert$\,$\leq$\,10) that cannot be detected in photometry. Only one of their
frequencies (45.44$\pm$0.02\,d$^{-1}$) was detected in our analysis. It matches our $f_{27}$=45.43672\,d$^{-1}$ frequency. 
\cite{Koen+2003b} provide some mode identifications but, as they discuss, there may be some doubt about the validity 
of the results. In particular they did not have enough data to avoid the one-day alias in the frequency determination and 
 the fast rotation of the star complicates the problem even further. 
%

\section{Conclusions}
\label{sect:conclusion}

We have presented and analysed high-precision photometric observations 
of $\beta$~Pictoris, covering the period from March to September 2017, using the  ASTEP 400 mm telescope installed at 
Concordia, Dome\,C in Antarctica. We detected 31 pulsation frequencies among which 28 are new. All of the detected frequencies were in the 
interval 34.76\,-\,75.68\,d$^{-1}$. We found amplitude variation for one pulsation mode, but there was no significant phase change in any
detected frequencies. 

In this present study, we focussed only on the detection of the pulsation frequencies of the star. A future work will be devoted to the 
identification of these pulsation modes by combining photometric and radial velocity data. Furthermore, the BRITE-Constellation 
mission \citep{Weiss+2014} has observed $\beta$~Pictoris continuously in two colours from November 4, 2016 to June 22, 2017; the 
corresponding results will be described in Zwintz et al. (in preparation).

The pulsation spectrum, obtained from the ASTEP data, already allows the calculation of a synthetic 
photometric stellar pulsation curve over a large observation period. This information is crucial to remove 
the $\delta$~Scuti pulsations signal from the $\beta$~Pictoris light curve to detect signatures of photometric events 
related either to the Hill sphere transit of $\beta$~Pictoris~b or to any other circumstellar material passing in front of the star.

\section*{Acknowledgments}
The field activities at Dome\,C benefit from the support of the French and Italian polar agencies IPEV and 
PNRA in the framework of the Concordia station programme. Part of the project has been supported by the Programme National de 
Plan\'etologie, the Lagrange Laboratory of the Observatoire de la C\^ote d'Azur, the National Research Agency (ANR-14-CE33-0018), and through 
the $\rm UCA^{JEDI}$ ``Investments in the Future'' project managed by the ANR (ANR-15-IDEX-01). 
KZ acknowledges support by the Austrian Fonds zur F\"orderung der wissenschaftlichen Forschung (FWF, project V431-NBL), the Austrian Space Application 
Programme (ASAP) of the Austrian Research Promotion Agency (FFG), and the Tiroler 
Wissenschaftsfonds (GZ: UNI-0404/1985; PI: K. Zwintz).  P.K. and J.J.W. acknowledge support from NASA NNX15AC89G 
and NNX15AD95G/NEXSS, NSF AST-1518332 and HST-GO-14621.  This work benefited from NASA's Nexus for Exoplanet 
System Science (NExSS) research coordination network sponsored by NASA's Science Mission Directorate. 



\bibliographystyle{aa}
\bibliography{astep_bp_pulsations.bbl}







\end{document}